\documentclass[preprintnumbers,superscriptaddress,nofootinbib]{revtex4-1}
\pdfoutput=1
\usepackage{graphicx}
\usepackage{amsmath,amssymb,enumerate}
\usepackage{slashed}
\usepackage{amsfonts}
\usepackage{geometry}
\usepackage{color}
\usepackage{appendix}
\usepackage{resizegather}
\usepackage{cancel}
\usepackage{bm,multirow}
\usepackage{subfig}
\definecolor{nicered}{rgb}{0.7,0.1,0.1}
\definecolor{nicegreen}{rgb}{0.1,0.5,0.1}

\usepackage{array}
\newcolumntype{L}[1]{>{\raggedright\let\newline\\\arraybackslash\hspace{0pt}}m{#1}}
\newcolumntype{C}[1]{>{\centering\let\newline\\\arraybackslash\hspace{0pt}}m{#1}}
\newcolumntype{R}[1]{>{\raggedleft\let\newline\\\arraybackslash\hspace{0pt}}m{#1}}

\usepackage[normalem]{ulem}
\usepackage{setspace}
\usepackage{multirow}

\def\wt{\widetilde}
\def\({\left(}
\def\){\right)}
\def\[{\left[}
\def\]{\right]}

\newcommand{\ba}{\begin{array}}
\newcommand{\ea}{\end{array}}
\newcommand{\bd}{\begin{displaymath}}
\newcommand{\ed}{\end{displaymath}}
\newcommand{\be}{\begin{equation}}
\makeatletter
\newcommand*{\rom}[1]{\expandafter\@slowromancap\romannumeral #1@}
\makeatother
\newcommand{\ee}{\end{equation}}
\def\bt{\begin{table}}
\def\et{\end{table}}
\def\bc{\begin{center}}
\def\ec{\end{center}}
\def\bi{\begin{itemize}}
\def\ei{\end{itemize}}
\def\bw{\begin{widetext}}
\def\ew{\end{widetext}}

\def\bea{\begin{eqnarray}}
\def\eea{\end{eqnarray}}
\def\beas{\begin{eqnarray*}}
\def\eeas{\end{eqnarray*}}

\def\N0{\widetilde{\chi}^0}

\def\a{\alpha}

\def\b{\beta}
\def\g{\gamma}
\def\d{\delta}

\def\G{\Gamma}

\def\l{\lambda}

\allowdisplaybreaks

\begin{document}

\title{Neutrino and Collider Implications of a Left-Right Extended Zee Model}

\author{Sarif Khan}
\email{sarifkhan@hri.res.in}
\affiliation{Harish-Chandra Research Institute, Jhunsi, Allahabad - 211019, India} 
\affiliation{Homi Bhabha National Institute, Training School Complex, Anushakti Nagar, Mumbai 400085,
India}
\author{Manimala Mitra}
\email{manimala@iopb.res.in}
\affiliation{Institute of Physics,Sachivalaya Marg, Sainik School Post, Bhubaneswar 751005, India}  
\affiliation{Homi Bhabha National Institute, Training School Complex, Anushakti Nagar, Mumbai 400085,
India}
\author{Ayon Patra} 
\email{ayon@okstate.edu}
\affiliation{Centre for High Energy Physics, Indian Institute of Science, Bangalore - 560012, India}

\preprint{%\begin{flushright}
IP/BBSR/2018-6}
 %\end{flushright}}

\begin{abstract}

We study a simple left-right symmetric (LRS) extension of the Zee model for neutrino mass generation. An extra $SU(2)_{L/R}$ singlet charged scalar helps in generating a loop-induced Majorana mass for neutrinos in this model. The right-handed neutrinos in this case are very light of the order of a few eV to a few MeV which makes this scenario quite different from other LRS models. We have analyzed the scalar potential and Higgs spectrum in detail, which also play an important role for the neutrino phenomenology. We identified the parameter regions in the model which satisfy the experimentally observed neutrino masses and mixings along with other experimental constraints. We have then studied the collider signatures of the charged scalar at $e^+e^-$ colliders with different benchmark points. It is possible to get a huge enhancement in the production cross-section of the charged scalar at lepton collider compared to the hadron colliders, resulting in a much stronger signal which can be easily observed at the upcoming ILC or CLIC experiments.

\end{abstract}

\maketitle

\section{Introduction} \label{intro}

The observation of neutrino oscillation leading to the realization that neutrinos are massive, is one of the biggest motivation for physics beyond the Standard Model (SM). A large number of models have been suggested to explain  neutrino masses and mixings either by the seesaw mechanism \cite{csaw} or through loop induced processes \cite{rad}. The Zee model \cite{zee} is one of the simplest such scenarios where neutrino masses are generated at  one-loop  by extending the SM scalar sector with an extra doublet and a charged singlet scalar field. The charged singlet scalar can mix with  other charged scalars while also having non-zero flavor violating couplings with  leptons, giving rise to neutrino masses at one-loop. Unfortunately the simplest form of the Zee model was shown to be ruled out by experimental neutrino data \cite{zee-out}.  However its extensions might still be viable. In this work we study an  extended  Zee model in a left-right symmetric (LRS) framework \cite{lr}. The model was proposed and studied in context of the LHC in \cite{pavel1} and the low energy flavor violating processes were discussed in \cite{pavel2}. In this work we examine its viability from neutrino oscillation data, study  the scalar potential in detail and derive charged  Higgs  spectrum, as well as analyze the possible electron-positron collider implications for the charged singlet Higgs boson.

Left-right symmetric (LRS) models are attractive extensions of the SM with the gauge group being extended to $SU(3)_C \times SU(2)_L \times SU(2)_R \times U(1)_{B-L}$. The parity symmetry is fundamentally conserved in these models which provides a natural solution to the strong CP problem \cite{scp} without introducing a global Peccei-Quinn symmetry. The parity symmetry is broken once the $SU(2)_R \times U(1)_{B-L}$ is spontaneously broken into $U(1)_Y$ at a scale $v_R$ much above the electroweak scale. Thus the observed parity violation in the SM can be easily understood. The gauge structure of LRS framework naturally requires the existence of right-handed neutrinos which can help generate light neutrino masses through seesaw mechanism. This usually requires the presence of an $SU(2)_R$ triplet scalar whose neutral component acquires a non-zero vacuum expectation value, leading to the right-handed symmetry breaking and the generation of Majorana masses for the right-handed neutrinos. The simplest LRS scenario, on the other hand, requires only an $SU(2)_R$ doublet scalar to achieve a consistent right-handed symmetry breaking but cannot generate light neutrino masses\footnote{These models need extra singlet fermions for neutrino mass generation.}. A simple LRS framework consisting of two doublets and a bidoublet scalar field, as will be considered here, can only generate a Dirac mass term for the neutrinos and the introduction of an extra charged singlet scalar is a very economical way to generate neutrino Majorana masses in such a scenario. Hence it is quite natural to extend the Zee model in a simple LRS framework to generate the neutrino masses and mixings.

There are several other advantages of  LRS extended Zee model. Firstly, since the neutrino Majorana masses are generated at one-loop, the right-handed neutrino masses also remain quite light ranging from a few MeV to a few eV. This is quite different from other LRS scenarios, where right-handed neutrinos are very heavy with masses proportional to the right-handed symmetry breaking scale (typically more than a few TeV). The presence of lighter right-handed neutrino states is a unique feature of this model. The recent results from the LSND \cite{Athanassopoulos:1995iw, Aguilar:2001ty} and MiniBooNE experiments  \cite{AguilarArevalo:2007it, AguilarArevalo:2010wv, Aguilar-Arevalo:2013pmq} hint at the existence of a light sterile neutrino with mass around a few eV. The LRS Zee model would be a prime candidate for explaining such a particle if these experimental results were to persist. Another important consequence of light right-handed neutrinos is the enhanced cross-section for the production of the $SU(2)_{L/R}$ singlet charged Higgs boson in this model, especially in the context of electron-positron colliders. The singlet charged Higgs bosons can be pair produced via a $t$-channel process. This process can either be mediated by a left-handed or a right-handed neutrino. The left-handed neutrino mediated processes suffer from extremely small couplings while the right-handed neutrino mediated processes (for models with heavy right-handed neutrinos) are suppressed by the large right-handed neutrino masses. The $t$-channel mediated charged Higgs pair-production cross-section thus remains extremely small for both these processes. Our scenario, with light MeV scale right-handed neutrinos, can alleviate this shortcoming and deliver large pair-production cross-section for the charged Higgs boson. Owing to the large couplings with the leptons, the charged singlet Higgs bosons can be copiously produced at lepton colliders, and  thus give rise to rich 
collider phenomenology. Since the singlet charged Higgs does not interact with the quarks of the SM, it has a limited discovery prospect in the hadronic colliders including the Large Hadron Collider (LHC).  A lepton collider, instead, is a perfect setup to test the singlet charged Higgs of this model. %Despite the limited capability of observing them at the LHC, the upcoming lepton colliders will be able to copiously produce . 

In this work, we pursue a detailed study of LRS extended Zee model by analyzing the neutrino mass and mixing constraints on the model parameters, taking into account  three generations of light neutrinos. We explicitly show the hierarchical structure of Dirac mass matrix. We also analyze the potential and evaluate the Higgs spectrum in detail. Furthermore, with the set of model parameters that satisfy neutrino oscillation measurements,  we carry out an in-depth analysis of the pair-production and decay of these charged scalars in the upcoming International Linear Collider (ILC) and Compact Linear Collider (CLIC) experiments. The final state of two opposite sign leptons and missing energy can be measured quite significantly over the SM background resulting in a possibility to observe such a process even with a very low luminosity $\mathcal{L} \sim 1-3 \, \rm{fb}^{-1}$ at these experiments. Therefore, even an early run of ILC/CLIC  can detect the presence of such a gauge singlet charged Higgs state. %\MM{Additionally, we fit  neutrino oscillation data and derive  constraints on  model parameters.  }

The rest of this paper is organized as follows. We discuss the model and the particle spectrum in Section.~\ref{model}. Following that, the pair-production of the charged Higgs and its detailed collider phenomenology is discussed in Section.~\ref{col}. We present our conclusions in Section.~\ref{conc}. 

\section{Model and Spectrum} \label{model}

LRS models are simple gauge extensions of the SM with the gauge group being $SU(3)_C \times SU(2)_L \times SU(2)_R \times U(1)_{B-L}$. The charge of a particle in this model is defined as 
\begin{equation}
\mathcal{Q}=I_{3L}+I_{3R}+\frac{B-L}{2},
\end{equation}
where $I_{3L/3R}$ is the third component of isospin under $SU(2)_{L/R}$ symmetry. The quarks and leptons consist of three generation of left-handed and right-handed doublet fields:
\begin{eqnarray}
\!\!Q_L\left (3,2, 1, \frac13 \right )\!\!&=&\!\!\left (\begin{array}{c}
u\\ d \end{array} \right )_L,~~
Q_R \left( 3,1, 2, \frac13 \right )\!=\!\left (\begin{array}{c}
u\\d \end{array} \right )_R ,\nonumber \\
l_L \left ( 1,2, 1, -1 \right )&=&\left (\begin{array}{c}
\nu\\ e\end{array}\right )_L,~~
l_R\left ( 1,1, 2, -1 \right )=\left (\begin{array}{c}
\nu \\ e \end{array}\right )_R,~~~
\end{eqnarray}
where the numbers in the brackets denote the quantum numbers under $SU(3)_C$, $SU(2)_L$, $SU(2)_R$, $U(1)_{B-L}$ gauge groups respectively. Here we see that the right-handed neutrinos are naturally present due to the gauge symmetry of the models.

The minimal Higgs sector, required for a consistent symmetry breaking mechanism and generation of quark and lepton masses and mixing angles, consists of
\begin{eqnarray}
H_R(1,1,2,1)&=&\left (\begin{array}{c}
H_R^+ \\H_R^0 \end{array} \right ),~~
H_L(1,2,1,1)=\left (\begin{array}{c}
H_L^+ \\H_L^0 \end{array} \right ),~~
\Phi(1,2,2,0)={\left (\begin{array}{cc}
\phi^{0}_1 & \phi^{+}_{2} \\ \phi^{-}_{1} & \phi^{0}_{2} \end{array} \right )},~~ \delta(1,1,1,2) = \delta^+ .~~~~~
\end{eqnarray}
The right-handed doublet field $H_R$ is required for breaking the $SU(2)_R \times U(1)_{B-L}$ into $U(1)_Y$ at some high scale to obtain the SM gauge symmetry at the electroweak (EW) scale. The $H_L$ doublet is required for preservation of the left-right symmetry. The bidoublet field $\Phi$ is responsible for generation of quark and charged lepton masses and Cabibbo-Kobayashi-Maskawa (CKM) mixing angles. The charged singlet field $\delta^{\pm}$ is needed for generation of neutrino masses through one-loop diagrams as will be discussed later in this section.

\noindent The Yukawa Lagrangian is given as:
\begin{eqnarray}
\mathcal{L}_Y&=&Y_{ij}^{q1}\overline{Q}_{Li}\Phi Q_{Rj}+Y_{ij}^{q2}\overline{Q}_{Li}\widetilde{\Phi}Q_{Rj}+Y_{ij}^{l1}\overline{l}_{Li}\Phi l_{Rj} + Y_{ij}^{l2}\overline{l}_{Li}\widetilde{\Phi}l_{Rj} + \l_{L_{ij}} l^T_{Li} i \tau_2 l_{Lj} \delta^+ + \l_{R_{ij}} l^T_{Ri} i \tau_2 l_{Rj} \delta^+ + H.C.~,~~
\label{eq:yuk}
\end{eqnarray}
where $Y$ and $\l$ are the Yukawa couplings and 
\begin{equation}
\widetilde{\Phi}=\tau_2\Phi^\ast\tau_2,%~~~~\widetilde H_{L/R} = i \tau_2 H^\ast _{L/R}.
\end{equation}
The structure of $\l_{L/R_{ij}}$ term is such that the only terms that will survive are the ones with $i \neq j$. This is exactly the same as in the Zee mechanism of neutrino mass generation. If we expand out any one of the terms involving $\d^+$ in the Yukawa Lagrangian we will get:
\begin{equation}
\mathcal{L} \supset \sum_{i \neq j} \nu_i e_j(\l_{ij}-\l_{ji}),
\end{equation}
where $\nu_i$ and $e_j$ are both in the flavor basis. Thus if we redefine the $\l$ matrix to  $\l'_{ij}=\l_{ij}-\l_{ji}$, then this new $\l'$ matrix is completely anti-symmetric and the Lagrangian terms can now be written as:
\begin{equation}
\mathcal{L} \supset \sum_{i,j} \nu_i e_j \l'_{ij}.
\end{equation}

%\section{Masses}
The Vacuum expectation values (VEVs) of the Higgs fields are given as:
\begin{equation}
\left<\phi_1^0\right>=v_1,~\left<\phi_2^0\right>=v_2,~\left<H_R^0\right>=v_R,~\left<H_L^0\right>=v_L,
\end{equation}
with the effective EW VEV given as $v_{EW}=\sqrt{v_1^2+v_2^2+v_L^2}$. Without loss of generality, one of the bidoublet VEVs can be chosen to be small. Also since $v_L$ does not contribute to the top mass, a large $v_L$ would automatically require a large top Yukawa coupling resulting in the theory being non-perturbative at quite low scales. The hierarchy in the VEVs thus has been chosen such that
\begin{equation}
v_R>>v_1>v_2,v_L.
\end{equation}

The gauge sector of the model consist of two charged $W_R^\pm$ and $W^\pm$ gauge bosons and three neutral bosons including the $Z_R$, $Z$ and the photon. The $W_R^\pm$ and the $Z_R$ bosons get their masses at the right-handed symmetry breaking scale and remain heavy while the others are the same as in the SM. The heavy gauge boson masses in this model are given as:
\begin{eqnarray}
M^2_{W_R^\pm} &&\simeq \frac{1}{2} g_R^2 (v_R^2+v_1^2+v_2^2), ~~~~M^2_{Z_R} \simeq \frac{1}{2} \left[(g_R^2+g_V^2) v_R^2+ \frac{g_R^4(v_1^2+v_2^2)+g_V^4 v_L^2}{g_R^2+g_V^2} \right],~~~~~~
\label{eq:ZR}
\end{eqnarray}
where $g_R$ and $g_V$ are the $SU(2)_R$ and $U(1)_{B-L}$ gauge couplings respectively. The left-handed (SM-like) gauge boson masses are given by their usual expressions with the effective $U(1)_Y$ gauge coupling $g_Y$ given as
\begin{equation}
g_Y=\frac{g_R g_V}{\sqrt{g_R^2+g_V^2}}.
\end{equation}

The scalar potential of this model is given as:
\begin{eqnarray}
V(\Delta,\Phi)&=& -\mu_{1}^2{\text{Tr}}\left(\Phi^\dagger \Phi \right)-\mu_2^2{\text{Tr}}\left[\widetilde{\Phi}\Phi^\dagger+\widetilde{\Phi}^\dagger\Phi\right]-\mu_3^2 H_R^\dagger H_R -  \mu_4^2 H_L^\dagger H_L- \mu_5^2 \delta^+ \delta^-+ \left(M_1 H_L^\dagger \Phi H_R \right.  \notag \\
&+& \left. M_2 H_L^\dagger \widetilde\Phi H_R + H.C.\right) + \l_1\left[{\text{Tr}}(\Phi^\dagger\Phi)\right]^2 + \l_2\left[\left\{{\text{Tr}}(\widetilde\Phi\Phi^\dagger)\right\}^2+\left\{{\text{Tr}}(\widetilde\Phi^\dagger\Phi)\right\}^2\right]+\l_3{\text{Tr}}(\widetilde\Phi\Phi^\dagger){\text{Tr}}(\widetilde\Phi^\dagger\Phi)\notag\\
&+&\l_4{\text{Tr}}(\Phi\Phi^\dagger)\left[{\text{Tr}}(\widetilde\Phi\Phi^\dagger)+{\text{Tr}}(\widetilde\Phi^\dagger\Phi)\right] + \left[ i \a_1 {\text{Tr}}(H_L^T \tau_2 \Phi H_R \delta^-) + i \a_2 {\text{Tr}}(H_L^T \tau_2 \widetilde\Phi H_R \delta^-)  + H.C.\right] \notag \notag\\
&+& \a_3(H_L^\dagger \Phi \Phi^\dagger H_L)  + \a_4\left[{\text{Tr}}(\widetilde\Phi\Phi^\dagger+\widetilde{\Phi}^\dagger\Phi)H_L^\dagger H_L\right]+\a_5{\text{Tr}}(\Phi\Phi^\dagger)H_L^\dagger H_L +\a_6 (H_R^\dagger \Phi^\dagger \Phi H_R)  \notag\\ 
&+&\a_7\left[{\text{Tr}}(\widetilde\Phi\Phi^\dagger+\widetilde{\Phi}^\dagger\Phi)H_R^\dagger H_R\right] + \a_8{\text{Tr}}(\Phi\Phi^\dagger)H_R^\dagger H_R + \b_1 (H_L^\dagger H_L)^2 + \b_2 (H_R^\dagger H_R)^2 +\b_3 (H_R^\dagger H_R)(H_L^\dagger H_L)  \notag \\
&+&\g_1 {\text{Tr}}(\Phi^\dagger \Phi) \delta^+ \delta^- + \g_2 {\text{Tr}}\left[\widetilde{\Phi}\Phi^\dagger+\widetilde{\Phi}^\dagger\Phi\right]\delta^+ \delta^- + \g_3 H_L^\dagger H_L \delta^+ \delta^- +\g_4 H_R^\dagger H_R  \delta^+ \delta^- + \g_5 (\delta{^+} \delta{^-})^2.
\label{eq:spot}
\end{eqnarray}

This gives four CP-even, two CP-odd and three charged Higgs boson states. Two CP-odd and two charged states are eaten up to give mass to the $Z_R,~Z,~W_R,W$ gauge bosons respectively. We will mainly focus our discussion on the charged Higgs sector, as that is the most important for the neutrino masses and the collider analysis which will be studied in this paper.

Minimizing the scalar potential of Eq.~\ref{eq:spot} we get four minimization conditions given as
\begin{eqnarray}
&&2 (\l_1 + 4 \l_2 + 2 \l_3) v_1 v_2^2  +  2 \l_4 v_2^3  + v_1 (\a_5 v_L^2 + \a_8 v_R^2 + 2 \l_1 v_1^2 - \mu_1^2) +  2 v_2 (\a_4 v_L^2 + \a_7 v_R^2 + 3 \l_4 v_1^2 - \mu_2^2) + M_2 v_L v_R=0, \notag \\
&&2 ( \l_1 v_2 + 3 \l_4 v_1) v_2^2+ v_2 \left\{ (\a_3 + \a_5) v_L^2 + (\a_6 + \a_8) v_R^2  + 2 (\l_1 + 4 \l_2 + 2 \l_3) v_1^2  - \mu_1^2 \right\} + M_1 v_L v_R \notag \\
&& + 2 v_1 ( \a_4 v_L^2+ \a_7 v_R^2 + \l_4 v_1^2 - \mu_2^2) =0, \notag \\
&& \left\{ 4 \a_4 v_1 v_2 + \a_5 v_1^2 + (\a_3 + \a_5) v_2^2  + 2 \b_1  v_L^2+ \b_3 v_R^2 - \mu_4^2\right\}v_L + M_2 v_1 v_R + M_1 v_2 v_R = 0, \notag \\
&& \left\{ 4 \a_7 v_1 v_2 + \a_8 v_1^2 + (\a_6 + \a_8) v_2^2 + 2  \b_2 v_R^2+ \b_3 v_L^2 - \mu_3^2 \right\} v_R + M_2 v_1 v_L + M_1 v_2 v_L =0.
\end{eqnarray}

Using these conditions along with the scalar potential, the charged Higgs mass-squared matrix in the gauge basis $({\phi_1^-}^*, \phi_2^+, H_R^+,H_L^+,\delta^+)$ is given as
\begin{equation}
M_{H^\pm}^2 = 
\begin{pmatrix}
M_{11} & M_{12} & \a_3 v_1 v_L - M_2 v_R & \a_6 v_2 v_R + M_1 v_L & -\a_2 v_L v_R \\
M_{12} & M_{22} & \a_3 v_2 v_L + M_1 v_R & \a_6 v_1 v_R - M_2 v_L & \a_1 v_L v_R \\
\a_3 v_1 v_L - M_2 v_R & \a_3 v_2 v_L + M_1 v_R & M_{33} & M_1 v_1 + M_2 v_2 & -(\a_1 v_2 + \a_2 v_1) v_R \\
\a_6 v_2 v_R + M_1 v_L & \a_6 v_1 v_R - M_2 v_L & M_1 v_1 + M_2 v_2 & M_{44} & (\a_1 v_1 + \a_2 v_2) v_L \\
-\a_2 v_L v_R & \a_1 v_L v_R & -(\a_1 v_2 + \a_2 v_1) v_R & (\a_1 v_1 + \a_2 v_2) v_L & M_{55}
\end{pmatrix},
\end{equation}
where 
\begin{eqnarray}
M_{11} &=& \left(-M_2 v_1 v_L v_R + M_1 v_2 v_L v_R + \a_3 v_1^2 v_L^2  + \a_6 v_2^2 v_R^2\right)/ (v_1^2 - v_2^2), \notag \\
M_{12} &=& (M_1 v_1 v_L v_R - M_2 v_2 v_L v_R + \a_3 v_1 v_2 v_L^2 + \a_6 v_1 v_2 v_R^2)/(v_1^2 - v_2^2),\notag \\
M_{22} &=& \left(-M_2 v_1 v_L v_R + M_1 v_2 v_L v_R + \a_3 v_2^2 v_L^2  + \a_6 v_1^2 v_R^2\right)/ (v_1^2 - v_2^2), \notag \\
M_{33} &=&  -\frac{1}{v_L} (M_2 v_1 v_R + M_1 v_2 v_R) +  \a_3 (v_2^2 - v_1^2), \notag \\
M_{44} &=& -\frac{1}{v_R} (M_2 v_1 v_L + M_1 v_2 v_L) +  \a_6 (v_2^2 - v_1^2), \notag \\
M_{55} &=&  \gamma_1 (v_1^2 + v_2^2)+ 4 \gamma_2  v_1 v_2 + \gamma_3 v_L^2 + \gamma_4 v_R^2 - \mu_5^2.
\end{eqnarray}
This $5\times 5$ charged Higgs mass-squared matrix can be diagonalized to obtain their mass eigenvalues as
\begin{equation}
M^2_{Diag} = V^{\dagger} M_{H^\pm}^2 V,
\label{eq:hdiag}
\end{equation}
where $M^2_{Diag}$ is the diagonalized charged Higgs boson mass-squared matrix and $V$ is the corresponding diagonalizing matrix. There are two zero eigenvalues corresponding to the two Goldstone bosons absorbed by the $W_R^\pm$ and $W^\pm$ bosons to give them mass. The Goldstone bosons primarily consist of $H_R^\pm$ and $\phi_1^\pm$ states respectively as their corresponding doublet neutral fields get the large non-zero VEVs. The other three eigenstates give the three physical charged Higgses and are  linear combinations of $\phi_2^\pm$, $H_L^\pm$ and $\d^\pm$. Flavor constraints, such as, $K^0-\bar{K}^0$ and $B^0-\bar{B}^0$ mixings require the neutral component of the bidoublet field $\phi_2^0$ mass to be heavier than 15 TeV~\cite{Zhang:2007da}, forcing its charged counterpart to be very massive as well. So $\d^\pm$ can primarily mix only with $H_L^\pm$ as $\phi_2^\pm$ is effectively decoupled owing to its large mass. We will consider two scenarios for our analysis. One where the lightest charged Higgs consists almost entirely of the charged singlet field $\d^\pm$ and another where the lightest physical state is almost equal admixture of $\d^\pm$ and $H_L^\pm$. In Tab.~\ref{tab:chig} we provide four benchmark points for the lightest charged Higgs boson $H_1^\pm$, two for the minimal mixing and two for the maximal mixing scenarios respectively.    
\begin{table}[h!]
\begin{center}
\begin{tabular}{C{1.5cm}|C{6.5cm}} \hline \hline
Mass & Composition \\ \hline
473.32 & $0.002 \phi_2^+ + 0.999 \d^+$ \\ \hline
1000.7 & $0.002 \phi_2^+ + 0.999 \d^+$ \\ \hline
432.58 & 0.03 ${\phi_1{^-}}^\ast-0.006 \phi_2^+ +0.72H_L^+ + 0.69 \d^+$ \\ \hline
1000.9 & 0.03 ${\phi_1{^-}}^\ast-0.006 \phi_2^+ +0.76H_L^+ + 0.65 \d^+$ \\ \hline \hline
\end{tabular}
\end{center}
\caption{Lightest charged Higgs boson $H^{\pm}_1$ eigenstates. The first two points correspond to minimal mixing while the next two are for maximal mixing.}
\label{tab:chig}
\end{table}
We  also cross-check  the corresponding scalar and pseudo-scalar neutral Higgs bosons for the set of parameters that we use to generate  the above charged Higgs masses, given in Table.~\ref{tab:chig}. We ensure   that the lightest scalar Higgs boson mass is  125 GeV and the pseudo-scalar sector has two massless Goldstone bosons,  required to give masses to the $Z_R$ and $Z$ bosons.

The quark and lepton masses can be obtained from Eq.~\ref{eq:yuk} as:
\begin{eqnarray}
M_{u} &=& Y^{q1} v_1+Y^{q2} v_2,~~M_d=Y^{q1} v_2+Y^{q2} v_1,~~M_l = Y^{l1} v_2+Y^{l2}v_1,~~M_\nu^D = Y^{l1} v_1+Y^{l2} v_2.
\end{eqnarray}
One can perform a simple rotation of the neutral bidoublet fields to obtain two new scalar fields 
\begin{equation}
h_1^0 = \frac{v_1 \phi_1^0 + v_2 \phi_2^0}{\sqrt{v_1^2 + v_2^2}},~~~~ h_2^0 = \frac{v_2 \phi_1^0 - v_1 \phi_2^0}{\sqrt{v_1^2 + v_2^2}}.~~~~
\end{equation} 
In this rotated basis, only one of these new fields ($h_1^0$) gets a non-zero VEV. This along with a redefinition of the couplings gives 
\begin{equation}
\hspace*{-0.3cm}
{M_{u} = Y^q v'_1,~~M_d = \wt{Y}^q v'_1,~~M_l = \wt{Y}^l v'_1,~~M_\nu^D = Y^l v'_1,}
\label{eq:fmass}
\end{equation}
where $\left< h_1^0 \right> = v'_1 $ is the VEV in the redefined basis and 
\begin{eqnarray}
Y^q &=& \frac{1}{v'_1}\left( Y^{q1} v_1+Y^{q2} v_2 \right),~~\wt Y^q = \frac{1}{v'_1}\left( Y^{q1} v_2+Y^{q2} v_1 \right),\notag \\ \wt{Y}^l &=& \frac{1}{v'_1}\left(Y^{l1} v_2+Y^{l2}v_1\right),~~Y^l = \frac{1}{v'_1}\left(Y^{l1} v_1+Y^{l2}v_2\right).~~
\end{eqnarray}  
The $\delta^+$ field is responsible for producing the Majorana mass terms in the neutrino mass matrix which are given as \cite{pavel1}:
%\begin{widetext}
\begin{eqnarray}
\displaystyle {(M_\nu^L)}^{\alpha \gamma}&&=\frac{1}{4\pi^2}\lambda'_L{^{\alpha \beta}}m_{e_\beta}\sum_{i=1}^3 \text{Log}\left(\frac{M_{h_i}^2}{m_{e_\beta}^2}\right) \times V_{5i}\left [ (Y_l^\dagger)^{\beta \gamma}V_{2i}^*-(\wt Y_l^\dagger)^{\beta \gamma}V_{1i}^* \right ] \ + \ \alpha \leftrightarrow \gamma 
\,,\notag \\
\displaystyle {(M^R_\nu)}^{\alpha \gamma}&&=\frac{1}{4\pi^2}\lambda'_R{^{\alpha \beta}}m_{e_\beta}\sum_{i=1}^3\text{Log}\left(\frac{M_{h_i}^2}{m_{e_\beta}^2}\right) \times V_{5i}\left[(Y_l)^{\beta \gamma}V_{1i}^*-(\wt Y_l)^{\beta \gamma}V_{2i}^*\right] \ + \ \alpha \leftrightarrow \gamma \,. 
\label{RHM}
\end{eqnarray}
%\end{widetext}
Here $\alpha$, $\beta$ and $\gamma$ each run from $1-3$, $V_{ij}$ corresponds to the $ij$-th element of the charged Higgs boson mixing matrix $V$ defined in Eq.~\ref{eq:hdiag}, $M_{h_i}(i=1-3)$ is the mass of the charged Higgs boson eigenstates and $m_{e_{\alpha}}$ is the charged lepton mass with $\alpha = 1,2~\text{and}~3$ representing the electron, muon and tau respectively. The neutrino mass matrix would thus be a $6\times6$ matrix in the $(\nu_{L_i},\nu_{R_j})$ $(i,j=1-3)$ basis given as:
\begin{equation}
M_\nu = \begin{bmatrix}
M^L_\nu&M^D_\nu\\(M^D_\nu)^T&M^R_\nu
\end{bmatrix},
\end{equation}
where  $M_{\nu}^L$ and $M^R_{\nu}$ are generated at one-loop while $M^D_\nu$ is the neutrino Dirac mass term. With the seesaw approximation, the light neutrino mass matrix appears as a combination of Type-I and Type-II seesaw:
\begin{equation}
M_{\nu}=M^L_{\nu}-{M^D_{\nu}}^T {M^R_{\nu}}^{-1} {M^D_{\nu}}.
\label{t2t1}
\end{equation}

The redefined coupling $\wt Y_l$, which we have chosen to be diagonal, is entirely determined from the charged lepton masses as can be seen from Eq.~\ref{eq:fmass}. Similarly $Y^q$ (chosen to be diagonal) and $\wt Y^q$ can be determined from the up and down sector quark masses and CKM mixings. For the neutrino sector we first chose $Y_l$ to be zero to get the light neutrino masses and mixings from $M^L_\nu$ alone. This approach does not work as there are too few free parameters to fit the experimental neutrino data ($\lambda^{\prime}_L$ is anti-symmetric). We then considered the case with non-zero $Y_l$ while $\l'_L$ was chosen to be zero. The light neutrino masses in this case arises entirely from $M_\nu^D$ and $M_R$ similar to type-I seesaw mechanism:
\begin{equation}
M_{\nu}=-{M^D_{\nu}}^T {M^R_{\nu}}^{-1} {M^D_{\nu}}.
\label{t1}
\end{equation}

This gave us the correct experimentally observed masses and mixings for the light neutrino and hence this is the approach we have chosen for the neutrino sector \footnote{Even if we keep both $Y_l$ and $\lambda'_L$ to be non-zero, for which $M^L_{\nu} \neq 0$, the values of the elements of $\lambda'_L$ matrix satisfying the neutrino constraints turn out to be very small to have any observable consequences for our study.}.
\begin{center}
\begin{figure}[h!]
\centering
\includegraphics[height=3.0in, width=3.50in]{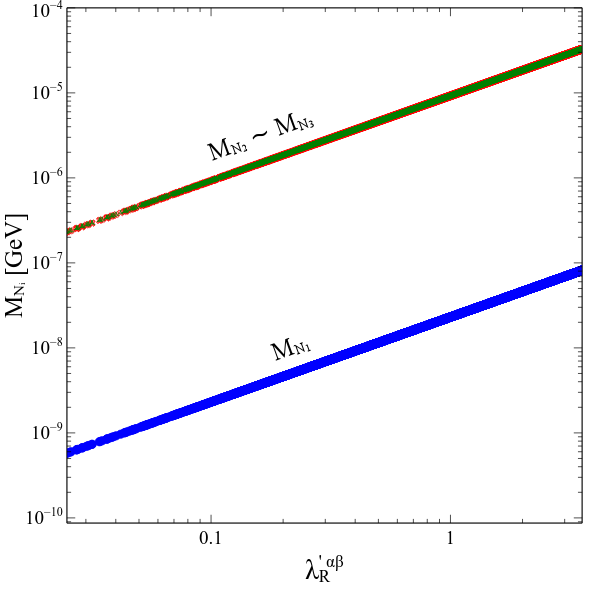}
\caption{Right-handed neutrino masses as a function of $\lambda'_R$ for minimal charged Higgs mixing.}
\label{fig:rhneumass}
\end{figure}
\end{center}
The right-handed neutrino masses in this scenario are generated at one loop and proportional to the square of the charged lepton Yukawa coupling  $\tilde{Y}_l$. Therefore,  right handed neutrino masses are quite small. As the other Yukawa coupling $Y_l$ is responsible for generating Dirac masses for the neutrinos, it  is  orders  of magnitude  smaller than $\tilde{Y}_l$ and hence does not have any impact on right-handed neutrino masses. We show the variation of the three right-handed neutrino masses $M_{N_{1,2,3}}$ with the Yukawa coupling $\lambda^{\prime}_R$  in  Fig.~\ref{fig:rhneumass}. As can be seen, that for $\lambda^{\prime}_R \sim 0.1-1$, the lightest right handed neutrino mass $M_{N_1}$ varies from  $3 \, \rm{eV}-30$ eV, while $M_{N_{2,3}}$ are in the sub-MeV scale.  In deriving this, we utilize Eq.~\ref{RHM}, where we diagonalize the right-handed Majorana mass matrix $M_\nu^R$. The charged Higgs boson masses and mixings used to obtain the neutrino Majorana masses are the ones corresponding to the first benchmark point in Tab.~\ref{tab:chig}. We have also provided these charged Higgs boson masses and mixings in details in Appendix A. This is quite different from other left-right symmetric models where the right-handed neutrino is naturally heavy as its mass is proportional to the right-handed symmetry breaking scale. 

\begin{table}

\begin{center}

\begin{tabular}{||C{9.0cm}|} \hline

 7.03$\times 10^{-5}~\text{eV}^2$ $<\Delta m_{21}^2<$ 8.09$\times 10^{-5}~\text{eV}^2$\\ \hline

 2.407$\times 10^{-3}~\text{eV}^2$ $<\Delta m_{31}^2<$ 2.643$\times 10^{-3}~\text{eV}^2$\\ \hline

 $0.271<\sin^2{\theta_{12}}<0.345$ \\ \hline

 $0.385<\sin^2{\theta_{23}}<0.635$ \\ \hline

 $0.01934<\sin^2{\theta_{13}}<0.02392$ \\ \hline

 $U_{PMNS}$ $$ \begin{pmatrix}

0.800 \rightarrow 0.844 & 0.515 \rightarrow 0.581 & 0.139 \rightarrow 0.155 \\

0.229 \rightarrow 0.516 & 0.438 \rightarrow 0.699 & 0.614 \rightarrow 0.790 \\

0.249 \rightarrow 0.528 & 0.462 \rightarrow 0.715 & 0.595 \rightarrow 0.776\end{pmatrix} $$ \\ \hline

\end{tabular}
\caption{{Experimental $3\sigma$ ranges for light neutrino parameters. See  \cite{bari,nufit} for further details.}}

\label{tab:neut}

\end{center}

\end{table}

\begin{center}
\begin{figure}[h!]
\centering
\includegraphics[width=4.5in]{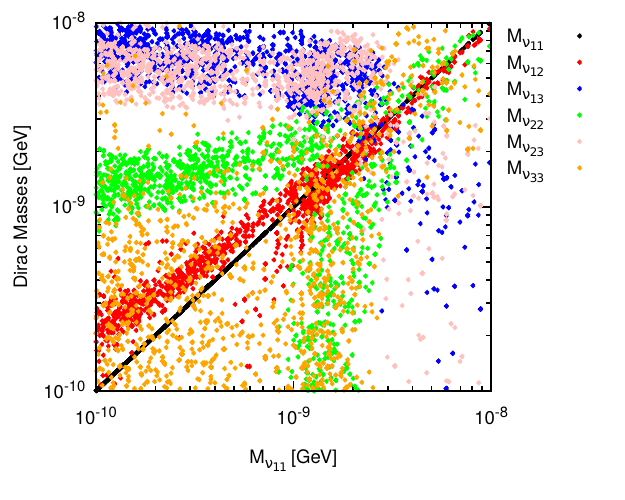}
\caption{Scatter plot of neutrino Dirac mass matrix elements $M^D_{\nu_{ij}}$ (denoted by $M_{\nu_{ij}}$ in the figure) satisfying the neutrino oscillation data in Table~\ref{tab:neut}.}
\label{fig:nyuk}
\end{figure}
\end{center}
\vspace*{-1cm}

For our subsequent analysis, we choose  $\lambda'_R \sim \mathcal{O}(1)$.  Since we use a type-I seesaw-like structure for the neutrino mass, the Dirac Yukawa couplings $Y^l$  in 
Eq.~\ref{eq:fmass}  are chosen accordingly to satisfy the correct neutrino oscillation parameters,  given in Table~\ref{tab:neut}.  As an illustrative example, we consider a normal hierarchy spectrum in light neutrino sector.  The allowed values for the elements of Dirac mass matrix $M^D_\nu$ are obtained by scanning over the allowed parameter space. We have varied the elements of $\l^{\prime}_R$ matrix between 0.5 to 1 keeping them very close to each other by allowing a spread of only 10\%. To generate $M^{R}_{\nu}$, we set the charged Higgs masses and mixings as given in Appendix A. Fig.~\ref{fig:nyuk} gives a scatter plot of the allowed neutrino Dirac masses (directly proportional to the Dirac Yukawa coupling $Y_l$) satisfying the experimental 3$\sigma$ ranges for the light neutrino parameters given in Tab.~\ref{tab:neut}. Here we plot the neutrino Dirac masses along the Y-axis with $M^D_{\nu_{11}}$ along the X-axis. This gives us an clear indication of the allowed values of the various terms in the $M^D_\nu$ matrix relative to each other. Note that the hierarchy between 
$(M^D_{\nu})_{11}, (M^D_{\nu})_{12},$ and $(M^D_{\nu})_{13}$ is clearly visible from the figure, with  $(13)$ element of the Dirac mass matrix allowed to take  highest values.  The hierarchy between $(13)$ and $(12)$ element is largest for lower $(M^D_{\nu})_{11}$ mass  $(M^D_{\nu})_{11} \sim 0.1 $ eV.   %Any possible hierarchy due to the experimental neutrino data is clearly visible in the $M_\nu^D$ sector. {\MM{\bf{Is this due to mass hierarchy or charged lepton mass??}}} %Since the lightest right-handed neutrino can be quite small, we also make sure that its mixing with the active neutrinos is small ($\sin \theta \lesssim 10^{-2}$).

As it is clear from the preceding discussion, in the present model we have an eV scale right-handed neutrino. Hence it may give the contribution to the relativistic degree of freedom (d.o.f) of the universe
if they equilibrate with the cosmic soup through their mixing with the active neutrinos.
Recently from Planck data there is a strong bound on the sum of the light degrees of freedom ({\it d.o.f}) which at $2\,\sigma$ gives $N_{\nu} < 3.2$\, and comes when we combine the $D/H$ ratio with the cosmic microwave background (CMB) baryon density \cite{Ade:2015xua, Cyburt:2015mya}.
However, the recent LSND \cite{Athanassopoulos:1995iw, Aguilar:2001ty} and MiniBooNE  \cite{AguilarArevalo:2007it, AguilarArevalo:2010wv, Aguilar-Arevalo:2013pmq}
data of electron excess in the antineutrino
mode requiress an eV scale sterile neutrino \cite{Abazajian:2012ys}.
The Reactor anomalies \cite{Mention:2011rk, Mueller:2011nm, Huber:2011wv, Abdurashitov:2005tb}
and the gallium experiments calibration data \cite{Bahcall:1994bq, Giunti:2006bj, Giunti:2010wz, Giunti:2010zu}
also hinted the presence of eV scale sterile neutrinos. 
Therefore, to go around the bound on the light relativistic {\it d.o.f}\,, a number of mechanisms have been suggested to overcome it.
Among them, the popular ones are as follows.
In \cite{Dasgupta:2013zpn, Hannestad:2013ana, Cherry:2016jol, Chu:2018gxk}, authors have used secret interactions where the sterile neutrinos are charged
under some hidden symmetry mediated by the light gauge boson, resulting in the mixing between active and sterile neutrinos being suppressed due to the
large thermal potential experienced by the sterile neutrinos. 
In \cite{Yaguna:2007wi}, they have shown that relativistic {\it d.o.f}
can be alleviated if the sterile neutrino is produced in a scenario where
the reheating temperature ($T_R$) is low, $T_{R} < 7$ MeV.
The authors of \cite{Saviano:2013ktj} have shown how to reduce
$N_{\nu}$ by studying the active-sterile flavor conversion.
In \cite{Ho:2012br}, they have used MeV dark matter to reduce $N_{\nu}$
with the help of $p$-wave annihilations. Ref.\,\cite{Giovannini:2002qw} discuss about the fact that without violating cosmology we can increase the relativistic {\it d.o.f} by reducing the neutron to proton ratio ($n/p$). A number of these possible resolutions can be applied for our model. For example, we can consider the existence of secret interactions with some hidden sector particles which would help lower the neutrino mixing between the left-handed and right-handed neutrinos. The effect of these interactions though would have ceased to exist at a much earlier time in the universe and today we will not be able to observe them anymore. Hence our current study would not be sensitive to them. Again, the other two right handed neutrinos are in MeV mass range and have a warm spectrum {\it i.e.} they are neither relativistic (which makes the problem with the cosmological structure formation \cite{Abazajian:2004zh}) nor non-relativistic. Extensive studies in the context of structure formation for such sub-MeV RH neutrinos are there in the literature \cite{Bode:2000gq, Hansen:2001zv, Boyarsky:2008xj, Boyanovsky:2010pw,Lovell, Boyanovsky:2010sv, Villaescusa, Merle:2013ibc, Abazajian:2012ys, Adhikari:2016bei}. Our model thus can be made consistent with the cosmological constraints but we have not considered them here as it is beyond the scope of this work.

\section{Experimental limits and Collider signature} \label{col}

In Table~\ref{tab:chig} we present a list of the various charged Higgs eigenstates that we consider in this study. We consider two cases with minimal mixings (thus consisting entirely of $\d^+$) and two with maximal mixing of $\d^+$ with $H_L^+$. For these benchmark points we study the pair production of charged Higgs states and their decay to a final state of two opposite sign charged leptons and two neutrinos. The most recent experimental bound on this process is from the ALTAS search \cite{Aaboud:2018jiw} of two opposite sign leptons and missing energy. They have put a bound of 500 GeV if the final state is coming from pair production of two sleptons.   The production cross-section of the charged Higgs at LHC   is however much lower for our model and even a 430 GeV charged Higgs is safe from the LHC bounds \footnote{For a set of loose cuts denoted by SF1 in \cite{Aaboud:2018jiw}, a production cross-section for $l^+ l^-  \cancel{E}_T$ greater than 2 fb is ruled out while we only get 0.23 fb for $M_{H^\pm}$ = 450 GeV with similar cuts.}. Therefore,  the benchmark points we have considered  are allowed by the experimental observations. 
\begin{figure}
\centering
\includegraphics[width=4.5in]{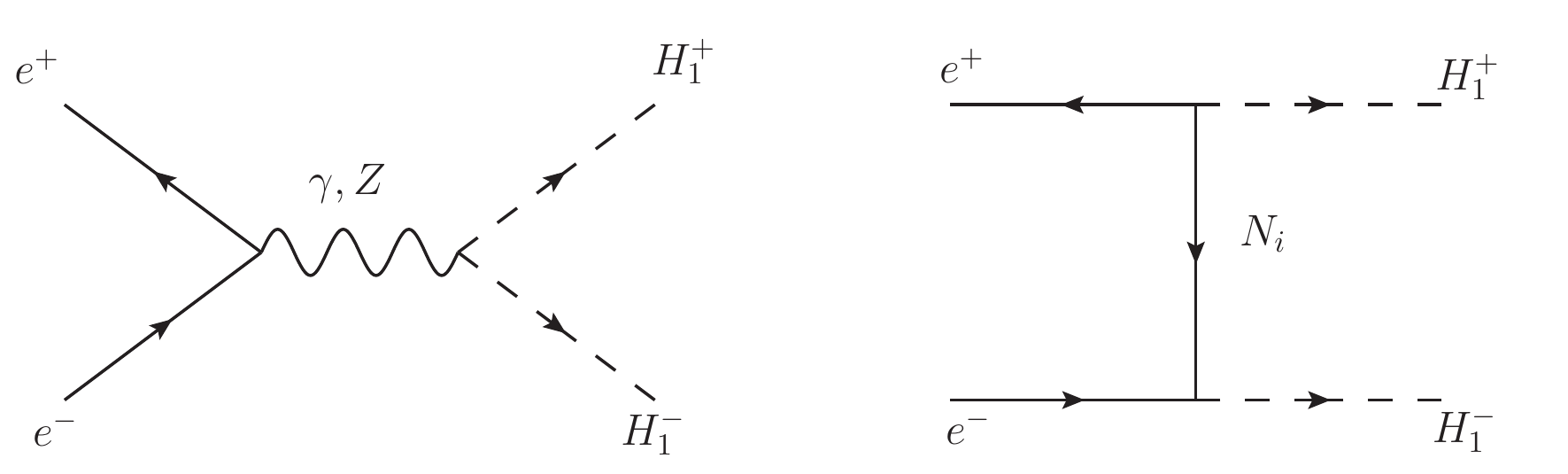}
\caption{Feynman diagram for the production  of $H^{+}_1H^{-}_1$
at $e^{+}e^{-}$ collider. The right panel diagram represents the contribution of the RH neutrinos in the pair-production process. }.  
\label{fig:feyn}
\end{figure}
The pair-production of the charged Higgs at LHC is through the s-channel process mediated by $\gamma$, $Z$ and $Z_R$ bosons which gives  small production cross-section.  In a lepton collider, on the other hand, there is an additional t-channel process mediated by the neutrinos as shown in Fig.~\ref{fig:feyn}. Owing to the large couplings of the charged singlet with the right-handed leptons and the small masses of the right-handed neutrinos in this model, this t-channel process will be the major pair-production channel. The masses of the right-handed neutrinos used in our analysis were taken as
\begin{equation}
M_{N_1} = 17~\text{eV},~M_{N_2} = 6.8~\text{KeV},~ M_{N_3} = 8.2~\text{KeV},
\end{equation}
for which the values of the Yukawa coupling $\lambda'_R \sim \mathcal{O}(1)$. We  thus study  the pair production of the charged Higgs at 1 TeV run of the International Linear Collider (ILC) \cite{ilc} and 3 TeV run by Compact Linear Collider (CLIC) \cite{clic}. 
\begin{figure}
\centering
\includegraphics[width=3.3in,height=3.0in]{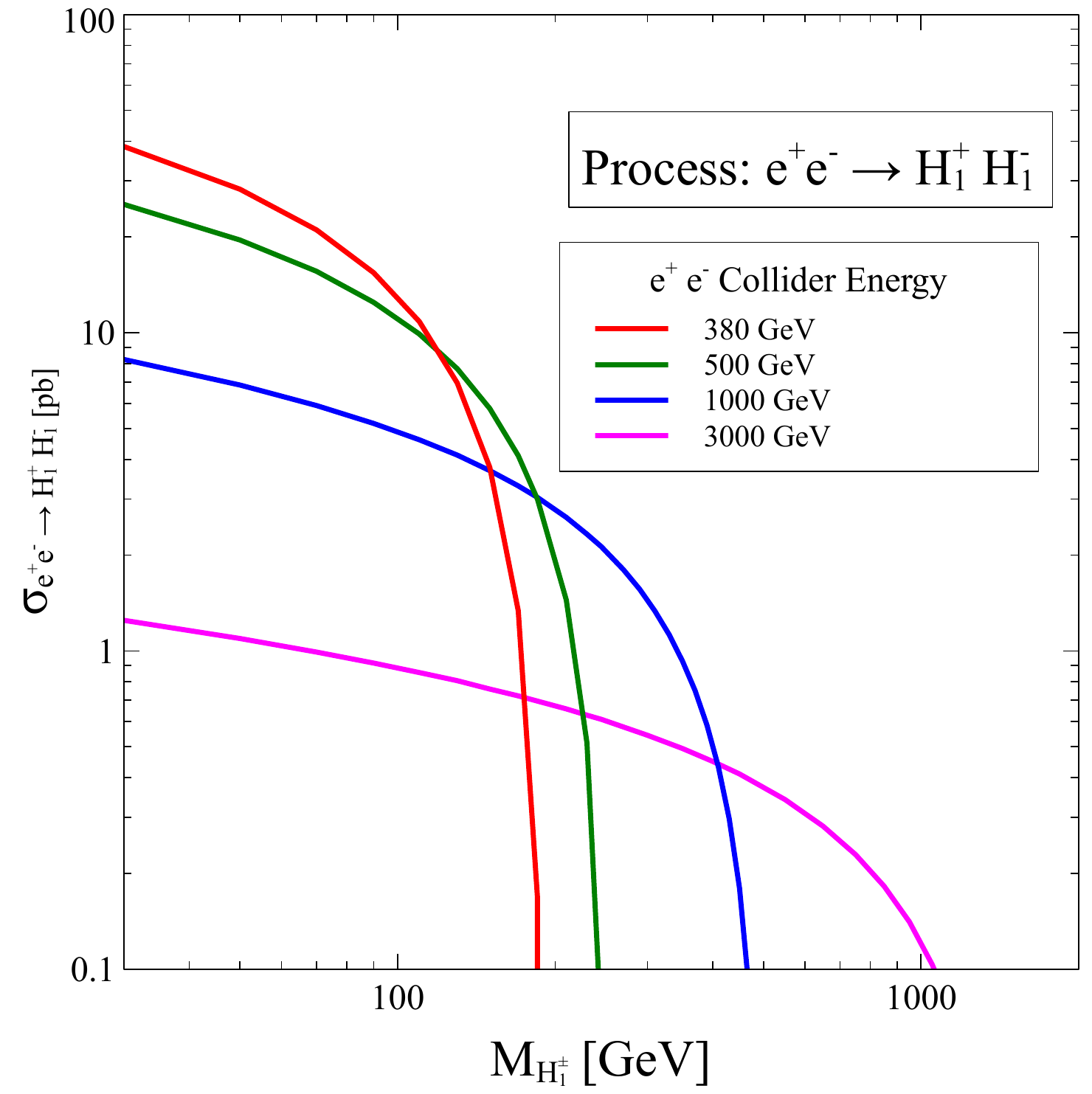}
\caption{Production cross section of $H^{+}_1 H^{-}_1$ at $e^{+}e^{-}$ collider for different center-of-mass energies.}      
\label{fig:prod}
\end{figure}
We  include the relevant vertices in  FeynRules  \cite{Alloul:2013bka}, and use  
MadGraph\cite{Alwall:2014hca} for event generation, Pythia
\cite{Sjostrand:2006za} for hadronization, and DelPhes
\cite{deFavereau:2013fsa} %, Selvaggi:2014mya, Mertens:2015kba} 
 for detector simulation.
Fig.~\ref{fig:prod} shows the  pair-production cross-section of the charged singlet Higgs as a function of its mass for four different center-of-mass (c.m.) energies at the lepton colliders. Here we  consider  the scenario of  minimal mixing of the singlet charged scalar for this figure. The charged Higgs  decays to a charged lepton and a right-handed neutrino and  gives rise to   a final state of dileptons with opposite charge ($l^{+}$ and $l^{-}$) and missing energy. Even the case where the charged Higgs is a mixture of $\d^\pm$ and $H_L^\pm$, this is the only kinematically allowed 2-body decay channel with its branching into 3-body decays being almost negligible. This is because $H_L$ does not couple to the quarks or leptons and its other physical states (the charged state with $H_L^\pm$ and $\d^\pm$ orthogonal to the one considered here and the CP-odd and CP-even neutral states coming from $H_L^0$) are much heavier. Schematically, the signal looks like
\begin{eqnarray}
e^+ e^{-} \rightarrow H^{+} H^{-} \rightarrow l^{+} l^{-} \cancel{E}_T+ X,
\end{eqnarray} 
where $l^{\pm}$ is either one of $e^{\pm}$, $\mu^{\pm}$ and $\tau^{\pm}$ or
combination of them. Inside the detector $\tau$ lepton will decay leptonically or hadronically and a small portion of it will give opposite sign dilepton and will increase the signal strength. As $\tau$ decays, eventually we get a final state signal which consist of opposite sign electron ($e^{\pm}$) or muon ($\mu^{\pm}$) or di-jet. For  simulation, we consider  Yukawa couplings $\lambda^{\prime}_R$, that  are allowed by  neutrino oscillation data.

%\section{Backgrounds}
Since we are interested in the opposite sign dilepton ($l^{+} l^{-}$) and missing
energy ($\cancel{E}_{T}$) signal in the final state, the corresponding SM
dominant backgrounds are as follows,
\begin{enumerate}
\item At the time of electron positron collision, opposite sign dilepton and missing energy can be produced as $e^+ e^- \rightarrow l^+ l^- Z\, (\rightarrow \nu_{l} \bar{\nu_{l}})$. This includes both the $ZZ$ and the virtual photon contribution. %{\bf{\MM{MM: virtual photon from which diagram?}}}}

%\item Background can arise from the Z Z final state production also,
%which is $e^+ e^- \rightarrow Z Z\, \rightarrow
%l^+ l^- \nu_{l}\, \bar{\nu_{l}}$.
\item Another dominant background is the $W^+ W^-$ pair production and its further leptonic decay. This can  mimic the signal as  
$ e^+ e^- \rightarrow W^{+} W^{-} \, \rightarrow l^+ l^- \nu_{l}\, \bar{\nu_{l}}$. 

\item Moreover, $t \bar{t}$ final state production and its subsequent decay
will also affect the signal as background in the following manner: $e^{+} e^{-} \rightarrow t(\rightarrow b\, l^{+} \nu_{l})\, \bar{t} (\rightarrow \bar{b}\, l^{-} \bar{\nu_{l}})$.

%\item Background which consist of a mono Z boson and jets which is $e^+ e^-
%\rightarrow Z+\, jets$.
\end{enumerate}  

%\section{Cuts used to remove the backgrounds}

\begin{figure}[t!]
\centering
\includegraphics[angle=0,height=5.5cm,width=5.7cm]{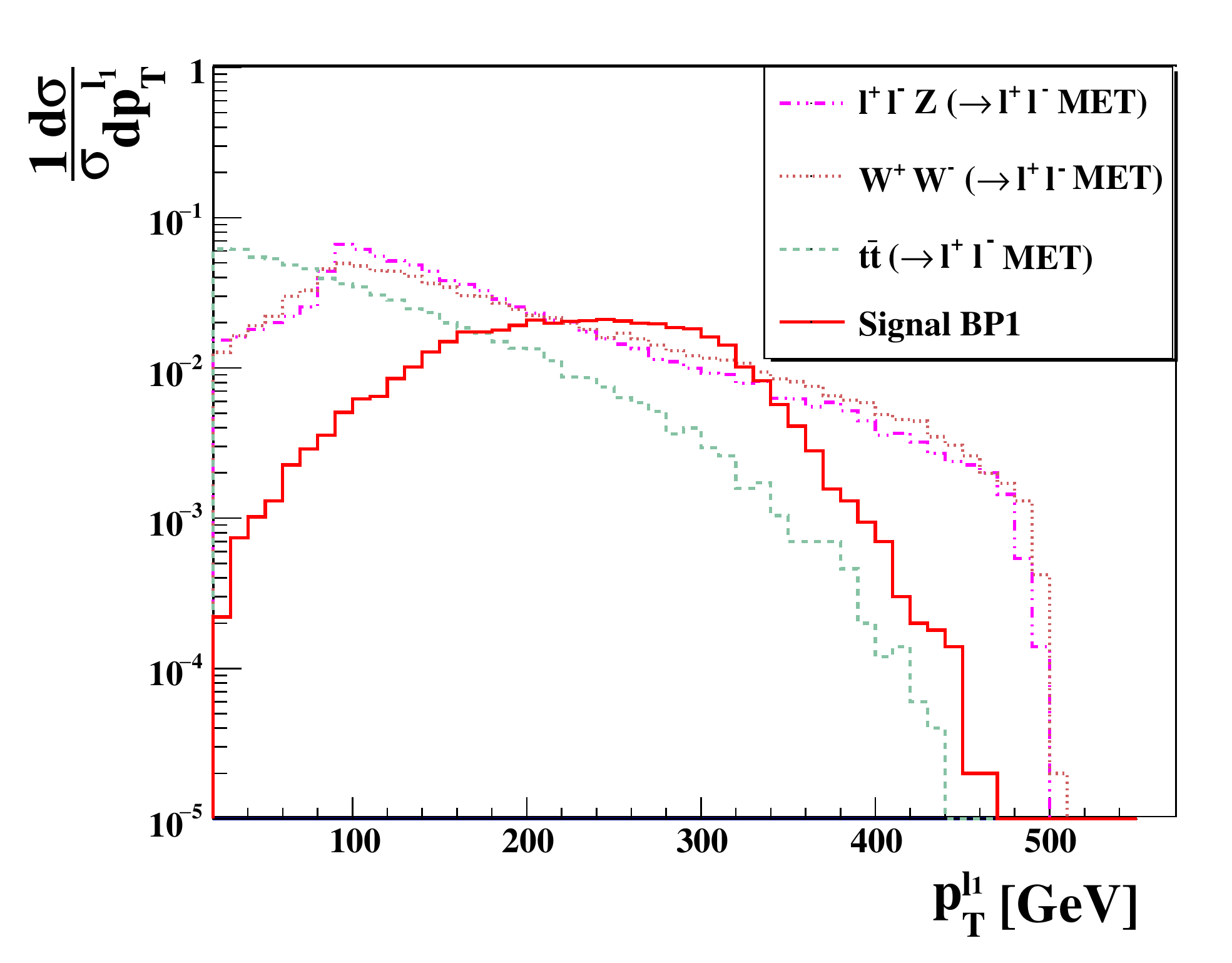}
\includegraphics[angle=0,height=5.5cm,width=5.7cm]{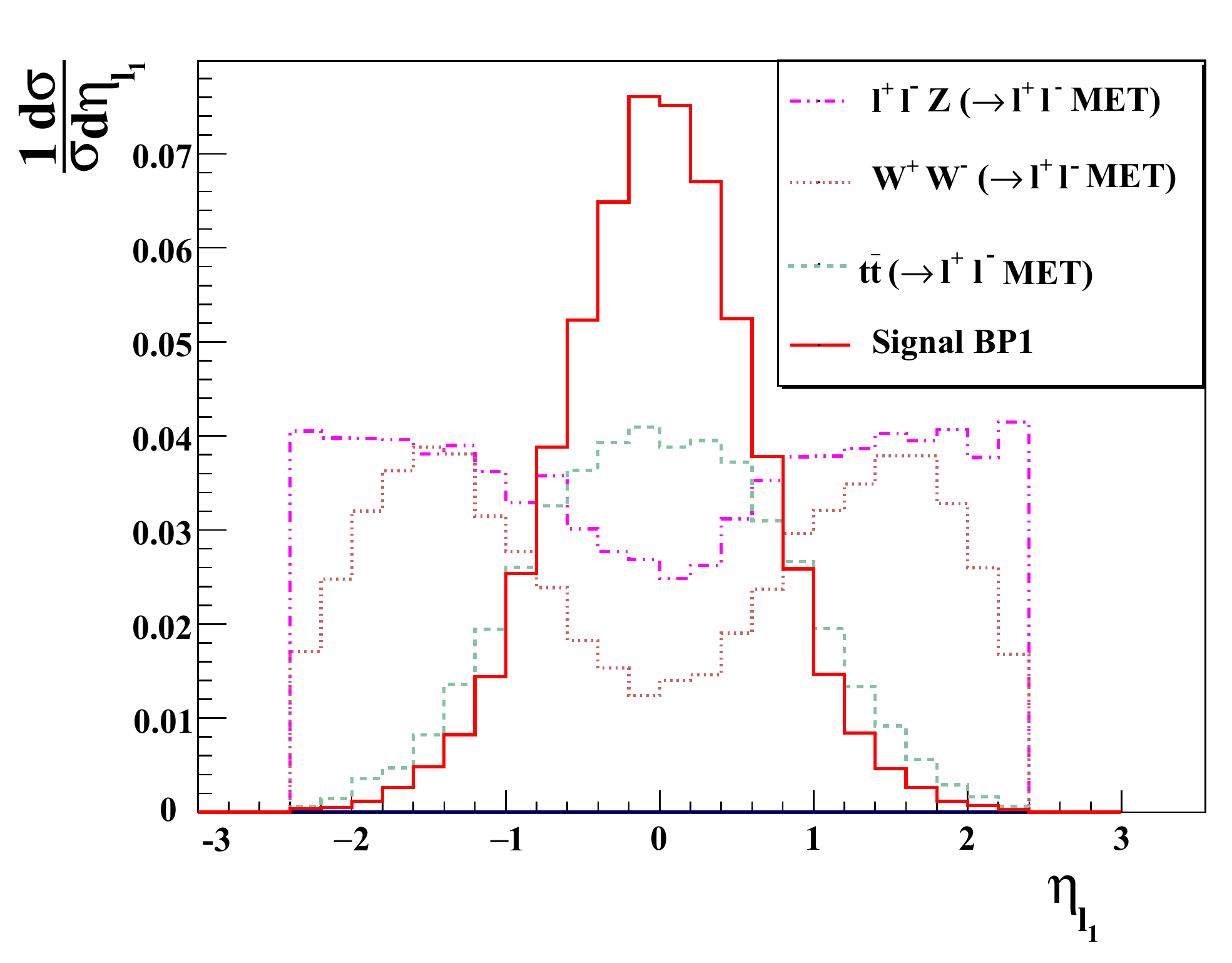}
\includegraphics[angle=0,height=5.5cm,width=5.7cm]{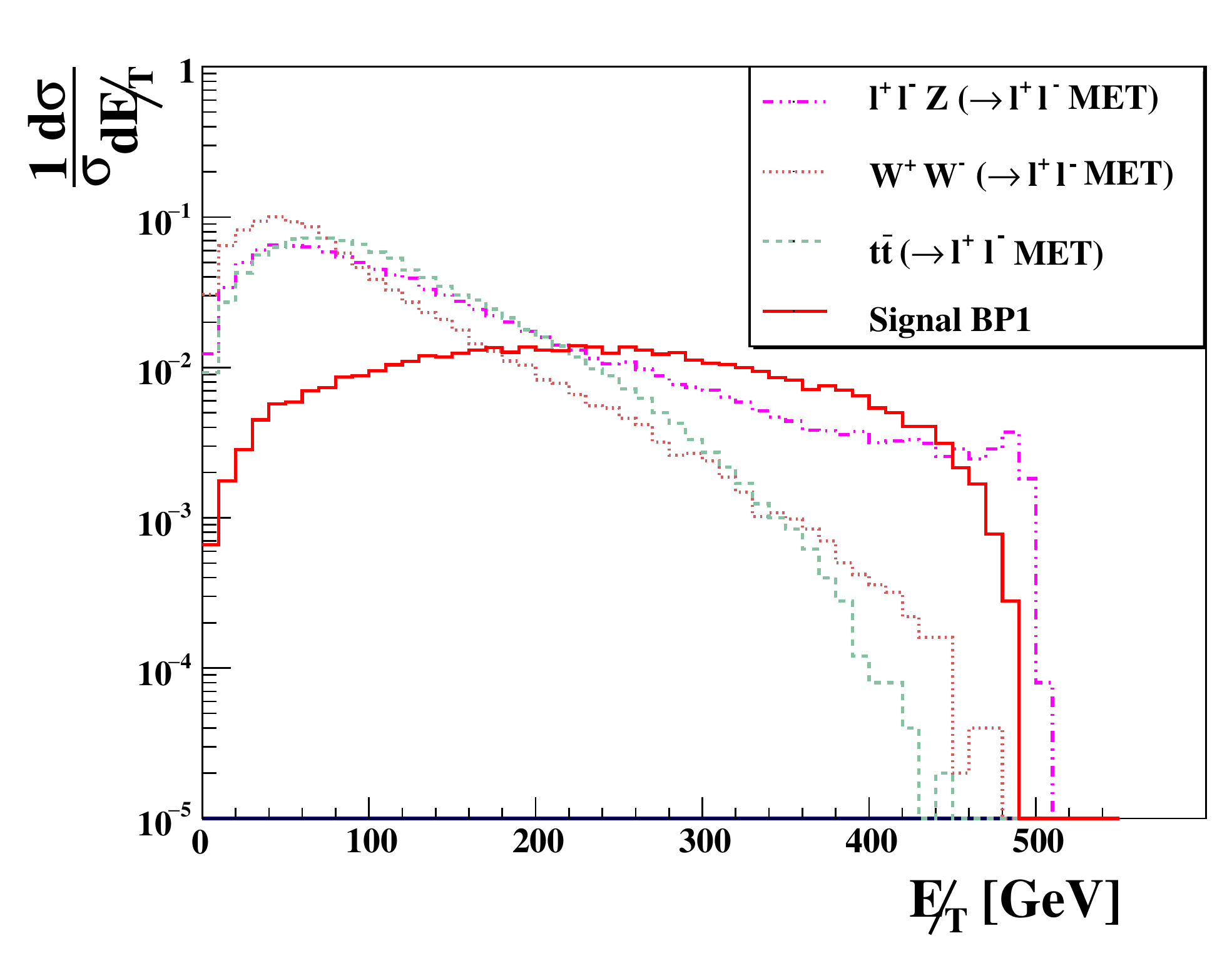}
\caption{Distribution of signal events and background processes for different kinematical variables. The plot on the left is the transverse momentum of the hardest lepton, the middle plot is for the pseudo-rapidity of the hardest lepton and the right plot is the  missing energy distribution.}      
\label{sig-bkg2}
\end{figure}

We do not put any veto on the light jet in our analysis. Additionally,  this is to note that the signal does not comprises of any $b$ jet. Therefore, a $b$-veto will  reduce the backgrounds, such as   $t\bar{t}$ production.
Depending on the various kinematical variables, there is a clear distinction between the signal and the backgrounds as can be seen clearly in Fig.\,\ref{sig-bkg2}. The leftmost plot in Fig.~\ref{sig-bkg2} shows the distribution of the transverse momentum of the hardest lepton $(p_T^{l_1})$, the one in the middle is its pseudo-rapidity distribution $(\eta_{l_1})$ and the rightmost plot shows the missing energy distribution of the signal and background events. Following these, we can select  appropriate cuts on different kinematical variables to  minimize the background,  while protecting the signal as much as possible. The details of the  cuts, that  we  use in our analysis are as follows:
\begin{enumerate}
\item[A0] We consider a signal in which final state contains two
opposite sign dilepton with missing energy {\it i.e., } $l^{+} l^{-} \cancel{E}_{T}$.
We implement a  minimum cut on the $p_T$ of the leptons which is
$p_{T,\,\, l}^{min} \geq 10$ GeV. We also implement  an upper limit on the pseudo-rapidity
which is $|\eta^{l}| < 2.5$. These cuts have been implemented  at the time of generating the partonic event samples.  % in the parton level.

\item[A1] We  select  our events which contains two opposite sign dilepton.

\item[A2] From the left panel of Fig.\,\ref{sig-bkg2} one can see that
if we use cut on the hardest lepton around 130 GeV then background can be reduced.
We therefore use  cut  on the $p_T$  of the hardest lepton, which is equal or greater than 130 GeV, $p_{T}^{l_1} \geq 130$ GeV and relatively softer cut on the second lepton which
is $p_{T}^{l_2} \geq 60$ GeV.  
 
\item[A3]  The  background from $ZZ$ pair production
can be safely removed by applying  $Z$-veto. We put a small
window on the dilepton invariant mass ($m_{ll}$) which is $|m_{ll} - 91.2| \leq 10$ GeV, and reject the events, that falls within this window.
%as clearly evident from the right panel of Fig.\,\ref{sig-bkg1}.

\item[A4] One of the background ($t\,\bar{t}$) contains $b$-jets in the final state. However,  
the signal of our interest doesn't have any $b$-jets.
Therefore we have used $b$-veto in the final state to reduce the
background without affecting the signal.  

\item[A5] From the middle  panel  of Fig.\,\ref{sig-bkg2}, it is evident  that
signal and backgrounds peak at different value
of the pseudo-rapidity of the leading  lepton. We
use tighter cut on $\eta^{l_1}$. We  reject   events which
have $|\eta^{l_1}| \geq 1 $. 

\item[A6] The RH neutrinos in our scenario are very light, as they have $\sim$ eV to MeV scale masses. The  decay of RH neutrinos can not happen inside
the detector. Hence,  they will be undetected and will give    missing energy.
We show the distribution of $\slashed{E}_T$ in the right panel of Fig.\,\ref{sig-bkg2}.   To reduce the background we also use 
cut on the missing energy, which is $\slashed{E}_{T} > 80$ GeV.
 This further enhances the signal to background ratio. 
\end{enumerate}

Using these above mentioned cuts,  we can reduce the background significantly while keeping the signal at a significant level. In Table~\ref{tab:1tev} and Table~\ref{tab:3tev}, we  show  the background cross-sections at 1 Tev ILC and 3 TeV CLIC experiments respectively,  after implementing all the  above-mentioned cuts.  The dominant background is $W^{\pm} W^{\mp}$ production, that has a cross-section $126.88$ fb 
(for 1 TeV ILC) at  partonic level. It is evident that the backgrounds  become quite small $\sigma \sim 7, 1$ fb for  ILC and CLIC respectively after the cuts. %and in both cases the dominant one is from the $W^+ W^-$ final state. 

%\begin{figure}[t]
%\centering
%\includegraphics[angle=0,height=7.5cm,width=8.0cm]{lep1_pt_new2.pdf}
%\includegraphics[angle=0,height=7.5cm,width=8.0cm]{inv_mass_new2.pdf}
%\caption{give caption}      
%\label{sig-bkg1}
%\end{figure}

%\section{Cut flow table for the Backgrounds for $e^{+} e^{-}$ Collider}
\onecolumngrid

\def\I{i}
\begin{center}
\begin{table}[ht!]
%\begin{center}
\begin{tabular}{||c|c||}
%\topline
\hline
\hline
\begin{tabular}{c|c}
    %\hline
    \multicolumn{2}{c}{SM Backgrounds at 1 TeV ILC}\\ 
    \hline
    Channels & Cross-section (fb) \\ 
    \hline
    $l^{+} l^{-} Z \,(\rightarrow \nu_{l}\bar{\nu_{l}}) $ & $18.68$ \\ 
    \hline
%    $Z(\rightarrow l^{+} l^{-})\, Z(\rightarrow \nu_{l} \bar{\nu_{l}})$ & $3.12$  \\
%    \hline
    $W^{+}(\rightarrow l^{+} \nu_{l})\, W^{-} (\rightarrow l^{-} \bar{\nu_{l}})$
    & $126.88$\\
    \hline
    $t (\rightarrow b l^{+} \nu_{l})\,\bar{t} (\rightarrow \bar{b} l^{-} \bar{\nu_{l}})$
    & $13.96$\\
    \hline
    Total Backgrounds & ~\\ 
\end{tabular}
&
\begin{tabular}{c|c|c|c|c|c}
    %\hline    
    \multicolumn{6}{c}{Effective Cross section after applying cuts (fb)}\\   
    \hline 
    A0\,+\,A1 & \,\,~~A2~~\,\, &\,\, ~~A3~~\,\, &\,\, ~~A4~~\,\, &\,\, ~~A5~~ &\,\,~~A6~~ \\
    \hline
    $10.79$ & $5.99$ & $5.54$ & $5.54$ & $2.30$ & $1.67$  \\
%    \hline
%    $1.33$ & $0.46$ & $0.03$ & $0.03$ & $0.02$ & $0.02$  \\
    \hline
    $52.72$ & $32.15$ & $32.15$ & $32.15$ & $12.44$ & $7.05$  \\
    \hline
    $3.10$ & $0.78$ & $0.78$ & $0.1$ & $0.08$ & $0.05$ \\
    \hline
%    19.66 & 5.37 & 3.59 & 0.50 & 0.12 \\
%    \hline
%    4.99 & 0.80 & 0.53 & 0.06 & 0.02  \\
%    \hline
    ~ & ~ & ~ & ~ & ~& 8.77 \\
\end{tabular}\\
\hline
\hline
\end{tabular}
%\bottomrule
%\hline
%\hline
\caption{Cut-flow table for the obtained cross-sections corresponding to the
different SM backgrounds. See the text for the details of the cuts   A0-A6. The c.m.energy is  
$\sqrt{s}=1 $ TeV, relevant for ILC.}
\label{tab:1tev}
%\end{center}
\end{table}
\end{center}

%%%%%%%%%%%%%%%%%%%%%%%%%%%%%%%%%%%

\vspace{-1.2cm}
%%%%%%%%%%%%%%%%%%%%%%%%%%%%%%%%%%%
\def\I{i}
\begin{center}
\begin{table}[h!]
%\begin{center}
\begin{tabular}{||c|c||}
%\topline
\hline
\hline
\begin{tabular}{c|c}
    %\hline
    \multicolumn{2}{c}{SM Backgrounds at $3$ TeV CLIC}\\ 
    \hline
    Channels & Cross-section (fb) \\ 
    \hline
    $l^{+} l^{-} Z \,(\rightarrow \nu_{l}\bar{\nu_{l}}) $ & $6.33$ \\ 
    \hline
%    $Z(\rightarrow l^{+} l^{-})\, Z(\rightarrow \nu_{l} \bar{\nu_{l}})$ & $0.13$  \\
%    \hline
    $W^{+}(\rightarrow l^{+} \nu_{l})\, W^{-} (\rightarrow l^{-} \bar{\nu_{l}})$
    & $13.85$\\
    \hline
    $t (\rightarrow b l^{+} \nu_{l})\,\bar{t} (\rightarrow \bar{b} l^{-} \bar{\nu_{l}})$
    & $1.76$\\
    \hline
    Total Backgrounds & ~\\ 
\end{tabular}
&
\begin{tabular}{c|c|c|c|c|c}
    %\hline    
    \multicolumn{6}{c}{Effective Cross section after applying cuts (fb)}\\   
    \hline 
    A0\,+\,A1 & \,\,~~A2~~\,\, &\,\, ~~A3~~\,\, &\,\, ~~A4~~\,\, &\,\, ~~A5~~ &\,\,~~A6~~ \\
    \hline
    $3.0$ & $2.89$ & $2.86$ & $2.86$ & $0.54$ & $0.44$  \\
    \hline
%    $0.05$ & $0.03$ & 0.002 & 0.002 & $3.0${\rm E-5} & $3.0${\rm E-5}  \\
%    \hline
    $5.45$ & $5.1$ & $5.1$ & $5.1$ & $1.34$ & $1.13$  \\
    \hline
    $0.05$ & $0.02$ & $0.02$ & 0.005 & 0.002 & 0.002 \\
    \hline
%    19.66 & 5.37 & 3.59 & 0.50 & 0.12 \\
%    \hline
%    4.99 & 0.80 & 0.53 & 0.06 & 0.02  \\
%    \hline
    ~ & ~ & ~ & ~ & ~& 1.57 \\
\end{tabular}\\
\hline
\hline
\end{tabular}
%\bottomrule
%\hline
%\hline
\caption{Cut-flow table for the obtained cross-sections corresponding to the
SM backgrounds. The details of the cuts A0-A6 are mentioned in the text. We perform the simulation for 
$3$ TeV CLIC.}
\label{tab:3tev}
%\end{center}
\end{table}
\end{center}
%%%%%%%%%%%%%%%%%%%%%%%%%%%%%%%%%%%

The signal cross-sections and their statistical significance over the background are given in Tab.~\ref{tab:signal} for the chosen benchmark points. Clearly the case with no mixing  in the Higgs state gives a much larger cross-section. This is because the $\d^\pm l_R^\mp \nu_R$ vertex is primarily responsible for the charged Higgs pair-production. The mixing of $\d^\pm$ with $H_L^\pm$ will only introduce an extra factor of $\cos^4 \theta$ in the pair-production cross-section, where $\theta$ is the mixing angle, resulting in a decrease in the cross-section. As evident the cross-section is enormous in the lepton collider.  As an illustrative example, for  a 1 TeV charged Higgs $H^{\pm}_1$, the partonic cross-section 
is $\sigma \sim 100$ fb. After the cuts, the cross-section reduces to $\sigma \sim 27$ fb. This is order of magnitude larger than the after-cut background cross-section. We compute the    statistical significance ($\mathcal{S}$) of signal over background  using the following expression,
\begin{equation}
\mathcal{S} = \sqrt{2 \times \left[ (s+b) {\rm ln}(1 + \frac{s}{b}) - s \right]}.
\end{equation}
 In the above, $s$ and $b$ denote the signal and background events. The significance has been shown in Tab.~\ref{tab:signal}.
As expected the case with zero mixing has a much better significance of signal over background boosting its chances to be discovered even in the early run of the upcoming lepton colliders. In particular, we show that only 
$\mathcal{L}=$1 $\rm{fb}^{-1}$ luminosity is required in the  zero-mixing scenario to discover  charged Higgs $H^{\pm}_1$ with mass range $473$ GeV - 1 TeV. For the relatively less optimistic  scenario of half-mixing, 3 $\rm{fb}^{-1}$ will be required to claim discovery.
%%%%%%%%%%%%%%%%%%%%%%%%%%%%%%%%%%%%%%%%%%%%%%
%%%%%%%%%%%%%%%%%%%%%%%%%%%%%%%%%%%
\onecolumngrid

\def\I{i}
\begin{center}
\begin{table}[!ht]
%\begin{center}
\begin{tabular}{||c|c|c||}
%\topline
\hline
\hline
\begin{tabular}{C{0.7cm}|C{1.2cm}|C{1.3cm}|C{1.1cm}|C{1.2cm}}
    %\hline
    \multicolumn{4}{c}{Signal at $e^{+}e^{-}$ Collider}\\
    \hline
   ~~ & COM Energy & Mass (GeV) & Mixing & CS (fb) \\
\hline
    BP1 & 1 TeV & 473.32 & Zero & 192.67 \\
    \hline
    BP2 & 3 TeV & 1000.70 & Zero & 100.31 \\    
    \hline
    BP3 & 1 TeV & 432.58 & Half & 49.50 \\
    \hline
    BP4 & 3 TeV & 1000.92 & Half & 17.86 \\
    
\end{tabular}
&
\begin{tabular}{C{1.2cm}|C{0.9cm}|C{0.9cm}|C{0.9cm}|C{0.9cm}|C{0.9cm}|C{0.9cm}}
    %\hline    
    \multicolumn{6}{c}{Effective Cross section after cuts (fb)}\\
    \hline 
    {\multirow{2}{*}{A0+A1}} & {\multirow{2}{*}{A2}} & {\multirow{2}{*}{A3}} & {\multirow{2}{*}{A4}} &{\multirow{2}{*}{A5}} & {\multirow{2}{*}{A6}}\\
    & & & & & \\
    \hline
    79.75 & 62.13 & 62.02 & 62.02 & 57.78 & 53.63   \\
    \hline
    38.21 & 35.57 & 35.56 & 35.55 & 28.08 & 27.07   \\    
    \hline
    19.19 & 14.62 & 14.59 & 14.59 & 13.54 & 12.51   \\
    \hline
    6.83 & 6.33 & 6.33 & 6.33 & 5.08 & 4.99   \\
    
\end{tabular}
&
\begin{tabular}{C{1.7cm}|C{1.7cm}}
    \multicolumn{2}{c}{Stat Significance ($\mathcal{S}$)}\\
    \hline    
     {\multirow{2}{*}{$\mathcal{L} = 1~{\rm fb^{-1}}$}}& {\multirow{2}{*}{$\mathcal{L} = 3$ ${\rm fb^{-1}}$}}\\ 
     &   \\
     \hline
     11.73 & 20.32\\ 
    \hline
     10.78 & 18.67\\
    \hline
     3.56 & 6.174\\ 
    \hline
     2.96 & 5.13\\

\end{tabular}\\
\hline
\hline
\end{tabular}
%\bottomrule
%\hline
%\hline
\caption{Cut-flow table of signal cross section at 1 TeV ILC and 3 TeV CLIC  after applying the different cuts.  We also show the statistical significance over the background.}
\label{tab:signal}
\end{table}
\end{center}

\section{Conclusion} \label{conc} 

In this work, we have studied left-right symmetric extension of Zee model. The model  has  very different characteristics features as compared to the minimal left-right symmetric model.  It is well known that the basic 
Zee model  is ruled out from light neutrino mass and mixing constraints. Going to the left-right symmetric framework, it is possible to evade the tension with the neutrino oscillation data. The model consists of three lighter right-handed neutrino states, that can have masses from MeV down to eV scale. Additionally, the model also contains an  additional charged scalar $\delta^{\pm}$. The charged scalar, due to its additional  interaction with  charged leptons and right-handed neutrinos, can be copiously produced at a lepton collider via the $t$-channel processes. 

We  discuss  light neutrino mass generation  in this model and fit the observed data.  The light neutrino mass matrix is a combination of both the Type-I and Type-II seesaw matrices. The  Type-II contribution and the right-handed neutrino mass matrix, that participates in Type-I seesaw,   are however generated through one loop process with the charged leptons and charged Higgs fields as mediators. %The right handed neutrino masses are  also generated through one loop process. 
We fit  the observed light neutrino mass square differences and the  PMNS mixing in this model, and derive  constraints  on model parameters.  With the set of parameters, that satisfy the neutrino mass constrains, we extensively analyze the charged Higgs phenomenology at 1 TeV ILC and 3 TeV CLIC.  Owing to the extra interaction of the charged Higgs with the right-handed neutrinos and for moderately large Yukawa couplings, the cross-section at $e^+e^-$ collider is enormous, as compared to the LHC. We find that in the most optimistic scenario, where the lighter charged Higgs state $H^{\pm}_1$ is a pure charged scalar state $\delta^{\pm}$, the cross-section  for pair-production of charged Higgs can be $\sigma \sim \mathcal{O}(1)$ pb  for $M_{H^{\pm}} \sim 473$ GeV, and  c.m.energy $\sqrt{s}=1$ TeV.  For CLIC, that can operate with c.m.energy $\sqrt{s}=3$ TeV, the charged Higgs of mass 1 TeV is also accessible ($\sigma \sim 100 $ fb for pair-production).

We consider the subsequent decay of the charged Higgs into  a lepton and a neutrino, that is the only possible channel for this model.  This leads to the  final states $l^{+}l^{-} + \slashed{E}_T$, that we analyze  in detail, taking into account detector simulation. We show that a discovery of the charged Higgs of mass in between 473-1000 GeV   in the di-lepton + $\slashed{E}_T$ will require only 1-3 $\rm{fb}^{-1}$ integrated luminosity at an $e^+e^-$ collider, operating with c.m.energy $\sqrt{s}=1, 3$ TeV. Therefore, this model can most economically be tested at the very early run of ILC or CLIC.

\begin{acknowledgements}

M.M. would like to acknowledge the DST-INSPIRE research grant IFA14-PH-99  and hospitality of CHEP, IISc, Bengaluru, where  part of the discussion has been carried out. A.P. is supported by the SERB National Postdoctoral Fellowship [PDF/2016/000202]. S.K. thanks Prof. Sandhya Choubey for discussions. S.K. also acknowledges the cluster computing facility at HRI (http://cluster.hri.res.in). 
S.K. would also like to thank the Department of Atomic Energy (DAE) Neutrino Project of Harish-Chandra Research Institute.
The authors would like to thank Dr. Arnab Dasgupta for very useful discussions at the early stage of this work.

\end{acknowledgements}
%%%%%%%%%%%%%%%%%%%%%%%%%%%%%%%%%%%
%%%%%%%%%%%%%%%%%%%%%%%%%%%%%%%%%%%%%%%%%%%%%%%%%%%%%%%%%%%%%%%%%%%%%%%%%%%%%%%%%%%%%%%
%\subsection{Statistical Significance}
%\onecolumngrid

%\def\I{i}
%\begin{center}
%\begin{table}[h!]
%%\begin{center}
%\begin{tabular}{||c|c|c||}
%%\topline
%\hline
%\hline
%\begin{tabular}{c|c|c|c|c}
    %\hline
%\multicolumn{4}{c}{~~~~~~~~~~~~~~~~~~~~~~Signal at $e^{+}e^{-}$ Collider}\\
%    \hline
%   ~~ & COM Energy & Mass (GeV) & Mixing & CS (fb) after cuts \\
%  \hline
%   BP1 & 1 TeV & 473.32 & Zero & 53.63\\
%    \hline
%    BP2 & 3 TeV & 1000.70 & Zero & 27.77\\   
%    \hline
%    BP3 & 1 TeV & 432.58 & Half & 12.51\\
%    \hline
%    BP4 & 3 TeV & 1000.92 & Half & 4.99\\ 
%  
%
%\end{tabular}
%&
%\begin{tabular}{c|c}
%    \multicolumn{2}{c}{Statitical Significance ($\mathcal{S}$)}\\
%    \hline    
%     $\mathcal{L} = 1$ ${\rm fb^{-1}}$& $\mathcal{L} = 3$ ${\rm fb^{-1}}$\\   
%     \hline
%     11.73 & 20.32\\ 
%    \hline
%     10.78 & 18.67\\
%    \hline
%     3.56 & 6.174\\ 
%    \hline
%     2.96 & 5.13\\
%
%\end{tabular}\\
%\hline
%\hline
%\end{tabular}
%%\bottomrule
%%\hline
%%\hline
%\caption{Statistical significance of the $l^{+} l^{-} \cancel{E}_T$
%signal for the 1 TeV and 3 TeV run of ILC and CLIC respectively. We have shown 
%the statistical significance for each
%run by considering 1 fb$^{-1}$ and 3 fb$^{-1}$ luminosity.}
%\label{tab6}
%%\end{center}
%\end{table}
%\end{center}
%\twocolumngrid
%%%%%%%%%%%%%%%%%%%%%%%%%%%%%%%%%%%%%%%%%%%%%%%%%%%%%%%%%%%%%%%%%%%%%%%%%%%%%%%%%%%%%%%
\numberwithin{equation}{section}

\begin{appendices}

\section{Charged Higgs boson eigenstates used for neutrino phenomenology}

Here we list the charged Higgs boson masses and mixings that have been  used for the neutrino phenomenology in our study. We consider that  the lightest charged Higgs boson $H^{\pm}_1$ has a mass around 473 GeV and is almost entirely consisting of the singlet charged Higgs field $\delta^{\pm}$. The charged Higgs boson states, after diagonalization,  consist of two Goldstone bosons $G_1^\pm$ and $G_2^\pm$ and three physical charged Higgs bosons with
\begin{equation}
M_{H_1^\pm} = 473.32~\text{GeV},~~M_{H_2^\pm} = 2534.94~\text{GeV},~~M_{H_3^\pm} = 15.95~\text{TeV}.
\end{equation}
The corresponding eigenstates can be identified as
\begin{equation}
\begin{pmatrix}
H_1^\pm\\H_2^\pm\\H_3^\pm\\G_1^\pm\\G_2^\pm
\end{pmatrix}
=
\begin{pmatrix}
0.0000127106 & 0.00225768 & 0.000109819 & 0.0000433475 &  0.999997 \\ -0.114973 & 0.000916758 & -0.993368 & 0.0000176018 & 0.000108482 \\ -0.000105399 & -0.999813 & -0.000910601 & -0.0191964 &  0.0022582 \\ -0.00574986 & -0.0191961 & 0.000665494 & 0.999799 & 4.70382 \times 10^{-16} \\ 0.993352 & -0.00011112 & -0.114971 & 0.00578718 & 
  2.76194 \times 10^{-18}
\end{pmatrix}
\begin{pmatrix}
\phi_1^\pm \\ \phi_2^\pm \\ H_L^\pm \\H_R^\pm \\ \delta^\pm
\end{pmatrix}.
\end{equation}
The $5\times5$ matrix in the above  is the charged Higgs boson rotation matrix $V$.

\end{appendices}

%%%%%%%%%%%%%%%%%%%%%%%%%%%%%%%%%%%%%%%%%%%%%%

\end{document}